\documentclass[twocolumn,pre,aps,nofootinbib]{revtex4}
\usepackage[latin9]{inputenc}
\usepackage{amsmath}
\usepackage{amssymb}
\usepackage{graphicx}

\makeatletter
\@ifundefined{textcolor}{}
{%
 \definecolor{BLACK}{gray}{0}
 \definecolor{WHITE}{gray}{1}
 \definecolor{RED}{rgb}{1,0,0}
 \definecolor{GREEN}{rgb}{0,1,0}
 \definecolor{BLUE}{rgb}{0,0,1}
 \definecolor{CYAN}{cmyk}{1,0,0,0}
 \definecolor{MAGENTA}{cmyk}{0,1,0,0}
 \definecolor{YELLOW}{cmyk}{0,0,1,0}
}

\usepackage{amsfonts}
\providecommand{\U}[1]{\protect \rule{.1in}{.1in}}

\makeatother

\begin{document}
\author{$^{1,2}$Miodrag L. Kuli\'{c}}
\title{High Pressure RTSC-Hydrides are Extreme Hard Type-II Superconducors}
\affiliation{$^{1}$Institute for Theoretical Physics, Goethe-University Frankfurt
am Main, Germany \\
$^{2}$Institute of Physics, 11080 Belgrade, Serbia}
\begin{abstract}
$\textrm{In}$\cite{Hirsch-1} the authors have called into question
the existence of the room-temperature superconductivity (RTSC) in
hydrides, as well as the electron-phonon interaction as the pairing
mecanism. Basically, in \cite{Hirsch-1} they assume, that these materials
are soft type-II superconductors with point-like defects as pinning
centers. We show here, that $\mathit{RTSC-hydrides}$ $\mathit{are}$
$\mathit{hard}$ $\mathit{type-II\textrm{ }superconductors}$, with
strong pinning centers in form of $\mathit{long}$ $\mathit{columnar}$
$\mathit{defects,}$with length $\mathit{L_{col}}$ much larger than
the superconducting coherence length $\xi(T)$ and penetration depth
$\lambda(T)$. It is shown that the $\mathit{elementary}$ $\mathit{pinning}$
$\mathit{energy}$ for the vortex lying along the long collumnar deffect
is given by $\mathrm{U_{col}\thickapprox U_{point}(L_{col}/\xi)}$.
At low temperatures it gives rise to high critical current density
$\mathrm{j_{c}\backsim10^{8}A/cm^{2}}$, which is of the order of
depairing current in hydrides. This theory predicts, that the temperature
width of the resistivity broadening (TRB), calculated in the Thinkam
theory, depends on $\mathrm{L_{col}}$, i.e. $\mathrm{\mathrm{\varDelta T}\sim\left(B\xi_{0}/L_{col}\right)^{1/2}}$.
Since $\mathrm{(\xi_{0}/L_{col})<10^{-4})\textrm{,}}$this width $\mathrm{\Delta T=\mathit{T_{c}(}1-t)}$
is at least $\mathit{100}$ times smaller in RTSC-hydrides, than that
predicted in \cite{Hirsch-1} for standard superconductors. However,
in the C-S-H superconductor with $\mathrm{T_{c}}$$\mathrm{=287}$
$\mathrm{K}$ the experimental TRB is almost field independent, which
probably needs a revision of the Tinkham theory. In an extreme hard
type-II superconductor irreversible phenomena are pronouced and the
penetration (trapping) of the field in and the magnetization hysteresis
are treated in the model of the Bean's critical state with long columnar
pinning centers. For instance, by assuming that the collective pinning
is realized in fields $\mathit{H\gg H_{c1}}$ one expects a significant
shift up of the irreversibility line $\mathit{B\sim(1-t)^{2}(L_{col}}/\xi).$
Due to the intrinsic hardness of the RTSC-hydrides they are promissing
candidate for producing high critical currents, especially if they
would be realized in systems which are in metastable state at low
pressure.
\end{abstract}
\date{\today}
\maketitle

\section{Introduction }

The first room-temperature superconductor is reported in 2015 in sulphur-hydrides
($H_{3}S$) under high pressure $P\approx150$ $GPa(\approx1.5$ $Mbar)$
\cite{Drozdov}. This opened a new frontier in physics and number
of other RTSC-hydrides were sinthesized. Let us mention some of them
with $\mathrm{T_{c}}$$\geq200$ $\mathrm{K}$, for instance $LaH_{10}$$\textrm{with}T_{c}\approx250$
\cite{LaH10}, $\mathrm{LaYH}$$_{x}$ with $\mathrm{T_{c}}$$\mathrm{=253}$
$\mathrm{K}$ \cite{Semenok}, C-S-H with $\mathrm{T_{c}}$$\mathrm{=287}$
$\mathrm{K}$ at $\mathrm{P}$$\approx267$. In that sense, the most
promissing candidate for the highest $\mathrm{T_{c}}$ is the metallic
hydrogen ($H$) which is the lightest atom in nature. One expects,
that in the metallic crystal state it could give as high as $T_{c}\sim600$
$K$, as it is theoretically predicted by Maksimov an Savrasov in
their seminal (but rarely cited!) paper in $\mathrm{2011}$ \cite{Maks-Savras-H}.
The pairing in the hydrogen-based systems is certainly due to the
strong electron-phonon interaction (EPI) via high-frequency $H$-phonons.
The pronounced isotope effect with $\alpha\sim0.3$ (in the pure EPI
theory for light atoms, such as $\mathit{H}$, even without effects
of Coulomb $\mu*$ term, $\alpha$ is smaller than the bookkeeping
value $\mathit{\mathrm{0.5}}$ (!)), is a clear evidence for the EPI
pairing in RTSC-hydides - see \cite{Gorkov-Kresin}, \cite{Kulic-rtsc}.
More on the pairing mechanism, especially on the role of the EPI,
in RTSC-hydride and other high-temperature superconductors such as
cuprates and Fe-based materials will be published elsewhere \cite{Kulic-rtsc}.

Many properties of RTSC-hydrides are unusual, since it turns out,
that they are extreme hard type-II superconductors with $\lambda\gg\xi$,
which can not be explained by the theory of soft superconductors.
Namely, in soft superconductors vortices are mainly pinned by randomly
distributed ``point'' defects \cite{Larkin},\cite{Civale-rev},
where the elementary pinning force of a ``point'' defect is due
the gain in the condensation energy when the vortex core is sitting
on the defect. The mechanism is simple, since in the ``point'' defect,
with the volume $V_{d}$, the superconductvity is completelly destroyed
(with $\varDelta=0\textrm{)}$ and the condensation energy $\mathrm{E_{loss}}$$\sim(\mathrm{H_{c}^{2}}/8\pi)$$\mathrm{V_{d}}$is
lost. So, if the part of the vortex (which core of the radius $\xi$
is in the normal state and the condensation energy is also lost) is
sitting on the ``point'' defect, then the loss of the condensation
energy is minimized. However, in order to maximize the pinning force
the radius of the defect should be $\sim\xi$ and $\mathrm{V_{d}}\sim\xi^{3}$(very
small defects with $\mathrm{d}$$\ll\xi$ are inapropriate since $\mathrm{V_{d}\sim\xi d^{2}}$
is small). This means that the gain of the energy is small due the
smallness of the ``point'' defect, while the Lorenz force acts on
the whole vortex line. As the result the critical current of the single
``point'' defect is small. On the other side, in the presence of
large densities of the ``point'' defects (which are uncorrelated)
two facts are against strong pinning. First, vortices can deform in
the zig-zag form, thus gaining pinning energy, but the zig-zag vortex
line increases its elestic energy. Additionally, each vortex interacts
with many (on the average uniformely) distributed defects, thus giving
on the average zero pinning force on it. However, there is still pinning
due to the local fluctuations of defects density but the pinning force
is small, since the maximum pinning force on the volume $\mathrm{V_{c}}$
is $\mathrm{F_{V_{c}}}$$\backsimeq f_{p}$$\sqrt{N}$. Here, $\mathrm{N=nV_{c}}$
is the number of the pinning centers in the volume $\mathrm{V_{c}}$and
$\mathrm{f_{_{p}}}$is the elementary pinning force. The critical
current density is given by $\mathrm{\mathrm{j_{c}}B=f_{p}\sqrt{N}/V_{c}}=\mathrm{f}$$_{p}(n/V_{c})^{1/2}$
\cite{Larkin}.

Therefore in order to obtain larger critical currant it is desirable
to have $\mathrm{long}$ ($\mathrm{of}$ $\mathrm{length}$ $\mathrm{L}$)
$\mathrm{defects}$ as pinning centers, with the large pinning volume
$\mathrm{V_{d}}$$=\pi\xi^{2}$$\mathrm{L_{col}}$ and the pinning
energy $\mathrm{U_{p}=u_{p}V_{col}}.$ In the case of the vortex (with
length $\mathrm{L_{v}>L_{col}}$sitting on such a defect the critical
current density is
\begin{equation}
j_{c}=\frac{c}{\varPhi_{0}}\times\frac{u_{p}\mathrm{V_{col}}}{\xi L_{v}}=\frac{c}{\varPhi_{0}}\times(\pi u_{p}\xi)\frac{L_{col}}{L_{v}}.\label{eq:j-c}
\end{equation}
It is clear, that in order to optimize the critical current density
it is desirable to have defects with the same length as the vortex
line, i.e $\mathrm{L_{col}\approx L_{v}}$ \cite{Kulic-CD} and with
the maximal possible density of the pinning energy $\mathrm{u_{p}.}$

Let us mention, that in the high-temperature superconductors the long
pinning defects were realized in the form of long columnar defects
by irradiating $\mathrm{YBa_{2}Cu_{3}O_{_{7}}}$ single crystals with
580 MeV $\mathrm{^{116}Sn^{30+}}$ ions, which produced long tracks
with the length $\mathrm{L_{col}\sim20\mu m}$ and the diameter 50$\textrm{Å}$.
In that case $\mathrm{j_{c}}$ reached $\mathrm{1.5}$$\times10^{7}$$\mathrm{A/cm^{2}}$
at $\mathrm{T=5K}$ and $\mathrm{10^{6}A/cm^{2}}$at $\mathrm{T=77K}$
\cite{Civale}, which are much higher than in the case of ``point''
defects. Also, the irreversible lines $\mathit{B}$$_{ir}(T)$ lie
much higher than in the case of the ``pont'' defects.

In the following we argue, that the vortex pinning by such long defects
is intrinsic property of the RTSC-hydrides. First, we introduce the
model of vortex pinning by long columnar defects where both, the core
and electromagnetic pinning, contribute almost equally to $\mathrm{u_{p}.}$
This is the most optimal situation for pinning. Next, the model is
applied to some ireversible prperties in the RTSC-hydrides, such as
temperature resistivity brodening in magnetic field, critical current
in the Bean-model as well as hysteresis in magnetization.

\section{Pinning of vortices by long columnar defects}

In Fig.1 are shown shematically columnar defects - blue coloured cylinders,
which strongly pin vortex lines in RTSC-hydrides, giving rise to large
magnetization hysterezis and critical current density in $\mathit{H}$$_{3}S$
see \cite{Troyan-1}. For simplicity we first analyze isotropis systems
\cite{Kulic-CD}. It turns out that for the optimal pinning superconductivity
is destroyed inside the defect, i.e. $\Delta=0$ and the radius of
the columnar defect should be small $\mathrm{\xi<R\ll}$$\lambda$
\cite{Kulic-CD}. In this case both mechanism of pinning, the core
and electromagnetic one, are operative and with almost the same energy
\cite{Kulic-CD}, $\mathrm{U_{p}=U_{core}+U_{em}}\approx2U_{cor}$,
where $\mathit{U}_{p}$$\approx2\pi\xi^{2}L_{col}(H_{c}^{2}/8\pi)=2L_{col}\Phi_{0}^{2}/(8\pi\lambda)^{2}$.
Since the pinning force (per unit vortex length $\mathit{L_{v}}$)
$f_{p}=$$\mathit{U_{p}/\xi L_{v}}$ in the critical state is balanced
with the Lorenz force (per unith length) $\mathit{f_{L}=j_{c}\Phi_{0}/c}$
then the critical current density is given by \cite{Kulic-CD}
\begin{equation}
j_{c}=\frac{c\Phi_{0}}{32\pi^{2}\lambda^{2}\xi}\left(\frac{L_{col}}{L_{v}}\right).\label{eq:jc-col}
\end{equation}
It is obvious that the critical current is maximized when the length
of columnar defect is equal to the vortex length, i.e. $\mathit{L_{col}=L_{v}}$.
In the case of anisotropic superconductors when the axis of the columnar
defect and the vortex are parallel to the crystal axis $\gamma$ and
the current flows along the $\beta$--axis (but $\alpha\neq$$\beta\neq\gamma\textrm{) it }$holds
\cite{Kulic-CD}
\begin{equation}
j_{c}^{\gamma,\beta}=\frac{c\Phi_{0}}{32\pi^{2}\lambda_{\alpha}\lambda_{\beta}\xi_{\alpha}}\left(\frac{L_{col}}{L_{v}}\right).\label{eq:jc-aniso}
\end{equation}
If one assumes that in RTSC-hydrides at $\mathit{T\ll T_{c}}$ one
has \cite{Drozdov} $\lambda\sim10^{3}\textrm{Å}$ and $\xi_{0}\sim20\textrm{Å}$
\cite{Drozdov} it follows that $\mathit{10^{7}(A/cm^{2})<j_{c}}$$\leq10^{8}A/cm^{2}$,
which is of the order of depairing current density. 

\section{Temperature broadening of resistivity in magnetic field}

The papers by Hirsh and Marsiglio \cite{Hirsch-1}, \cite{Hirsh-Meissner}
are stimulativ in understanding the physics of superconductivity in
the RTSC-hydrides and therefore deserve tribute, inspite of some controversing
claims. Namely, they were first to pay attention to some very pronounced
differences in properties of RTSC-hydrides and standard superconductors,
by including HTSC-cuprates and Fe-based superconductors. For instance
they point, that the temperature broadening of resistivity ($\mathit{TBR}$)
$\Delta T$ in several RTSC-hydrides is much smaller than in standard
superconductors by at least two order of magnitude. If one interprets
this fact in the framework of soft superconductors it is necessery
to invoke unrealistically high critical current density of the order
$\mathrm{10^{11}A/cm^{2}}$\cite{Hirsch-1}. Here, we argue that this
effect can be explained by assuming that RTSC-hydrides are hard type-II
superconductors with many intrinsic long pinning defects. Let us briefly
introduce the reader into the subject, which is based on the Tinkham
theory for $\mathit{TBR}$ \cite{Tinkham-1}.

\begin{figure}
\includegraphics[scale=0.25]{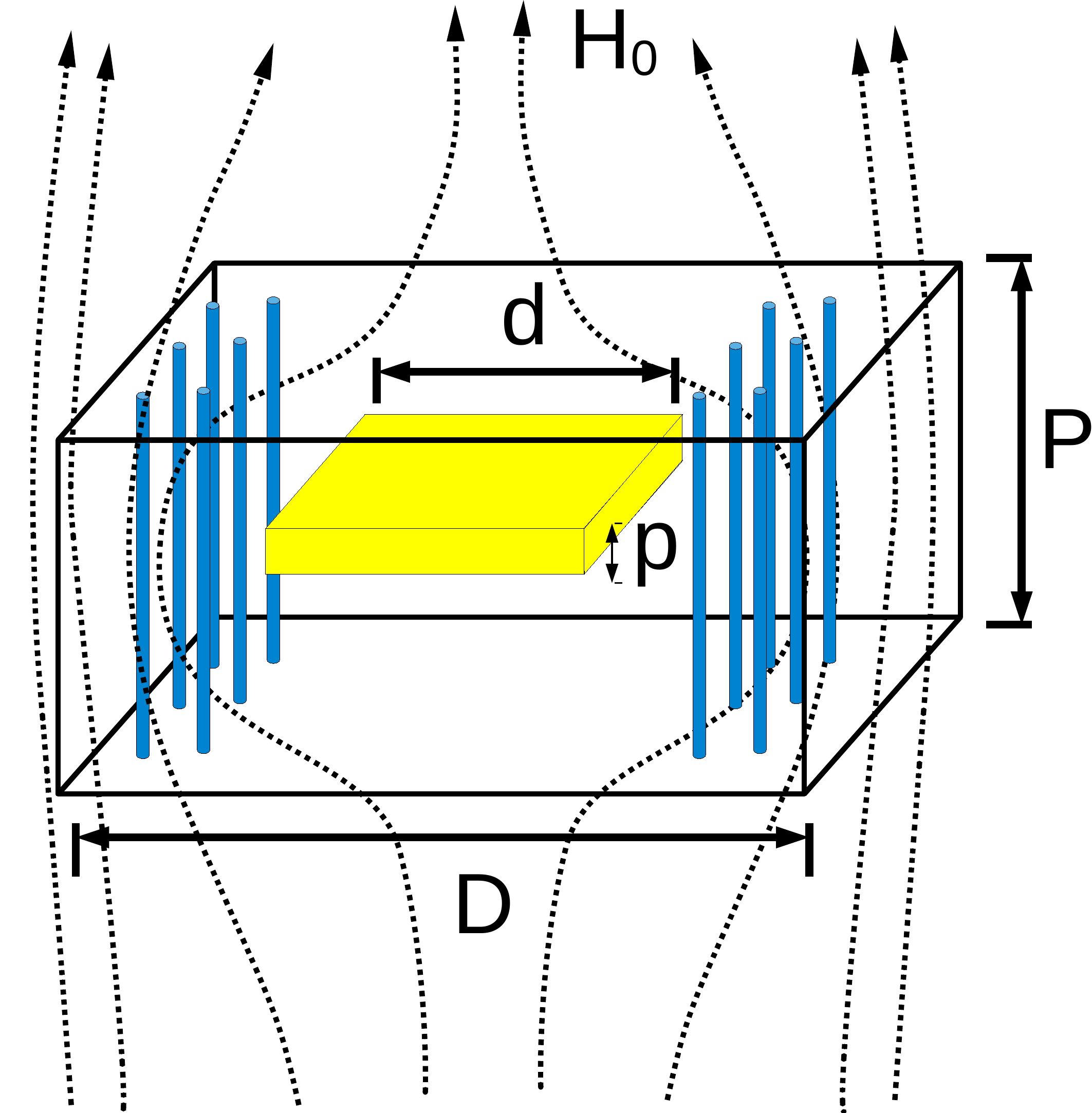}\caption{Shematic view of the experiment of the flux trapping and penetrating
in the $H_{3}S$ sample \cite{Troyan-1}: $\mathit{D=30\mu m;}P=5\mu m$.
The non-superconducting Sn foil ($\mathit{yelow}$) for detection
of the penetrated magnetic field: $\mathit{d=20\mu m;}p=2.6$$\mu m.$
Long columnar defects ($\mathit{blue}$ $\mathit{cylinders}$) strongly
pin and trapp vortices making huge magnetization hysterezis and critical
current density $\mathit{j_{c}\sim\Delta M}$.}
\end{figure}

Practically in all superconductors there are some defects which pin
vortices allowing a dissipationless current to flow even in the vortex
state. However, when the pinning energy is small there is a possibility
that vortices under temperature fluctuations jump from one center
to another thus giving rise to vortex motion and dissipation of energy.
This jumps are activation-like and proprtional to the escape (from
the pinning center) probability $\mathit{exp(-U_{p}/T)}$ - this is
so called flux creep \cite{Anderson}. Since for small defects$\mathit{U_{p}}$$\sim\xi^{3}$
then this energy barrier is small in superconductors with small $\xi$
what is the origin of prononced dissipation in HTSC-cuprates due to
small $\xi$. In that sense, a long pinning defect with $\mathit{U_{p}}$$\sim L_{col}\xi^{2}$
make this barier much higher, thus suppressing the dissipation defects.

\begin{figure}
\includegraphics[scale=0.75]{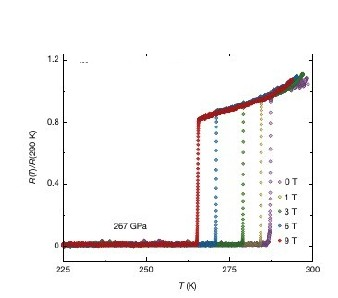}\caption{Low temperature electric resistance in magnetic field: H=0T, 1T, 3T,
6T and 9T ($\mathit{increasing}$ $\mathit{from}$ $\mathit{right}$
$\mathit{to}$ $\mathit{left}$) at 267 GPa - from \cite{Snider}.
Temperature broadening of resistivity is extremely small due to strong
pinning by columnar defects.}
\end{figure}

In magnetic fields much higher than the lower critical field $\mathit{H_{c1}\ll H<H_{c2}}$one
has $\mathit{B\approx H}$ and the vortex dinstance is given by $\mathit{a\approx l(\Phi_{0}/B)}\ll\lambda$
and the so called collective pinning of vortices should work (at least
it was useful for HTSC-cuprates) - where the bundlle of vortices are
pinned. In that case, one has $\mathit{U_{p}}$$\sim L_{coll}a^{2}$
(instead of $\mathit{U_{p}}$$\sim L_{col}\xi^{2}$ for the single
vortex pinning). The Thinkam developed the TRB theory which is based
on the Ambegaokar and Halperin theory for the thermaly activated phase
motion in Josephson junctions \cite{Ambegaokar}. As the result one
obtains the formula for the resistance in superconductors for small
currents \cite{Tinkham-1}

\begin{equation}
R/R_{N}=\left[I_{0}(\gamma_{0}/2)\right]^{-2},\gamma_{0}=U_{p}/T.\label{eq:R-Tink}
\end{equation}
$\mathit{R}$$_{N}$is the resistance of the normal state at $\mathit{T}_{c}$.
The modified Bessel function $\mathit{I_{0}}$gives the asymtotics
for $R/R_{n}\sim\gamma_{0}^{2}$ for $\gamma_{0}<1$ and $R/R_{n}\sim\gamma_{0}exp(-\gamma_{0})$
for $\gamma_{0}\gg1$ \cite{Tinkham-1}, where $\gamma_{0}$ is the
dimensionless pinning barrier with respect to temperature. In the
model with long columnar pinning centers and near $\mathit{T_{c}}$
one has 

\begin{equation}
\gamma_{0}^{col}=\beta_{K}\frac{L_{col}}{\xi_{0}}\frac{(1-t)^{2}}{b}(2\pi\xi_{0}^{2}\frac{j_{c}(0)\Phi_{0}}{cT_{c}}),\label{eq:gamma-0}
\end{equation}
where $\mathit{b=B/H_{c2}(0)}$ and $\beta_{K}\approx1$. In the case
of the ``point'' defects one has 
\begin{equation}
\gamma_{0}^{st}\approx\frac{(1-t)^{3/2}}{b}(2\pi\xi_{0}^{2}\frac{j_{c}(0)\Phi_{0}}{cT_{c}}),\label{eq:gamma-pd}
\end{equation}
where $\gamma_{0}^{st}$ is the prediction of the standard Thinkam
theory. From the above equations it is seen that relative barrier
in the long columnar case is larger by factor $\mathit{L_{col}/\xi(T)}$,
than the one for the ``point'' defects. Let us compare the prediction
of the satandard theory \cite{Tinkham-1} with the new one. Let us
assume, that we analyze the case when the resistance is measured at
the 10\% level, i.e. $\mathit{R/R_{N}=0.1}$ for the field $\mathit{B\approx1T}$
and $\mathit{H_{c2}(0)\approx100T},$i.e. $\mathit{b=10^{-2}}$. This
means that $\gamma_{0}^{col}=\gamma_{0}^{st}\equiv\gamma_{0}$. Since,
$\mathit{I_{0}(\gamma_{0}/2)=3}$ one has $\gamma_{0}\approx5$. We
assume the following realistic parameters for the RTSC-hydrides: $\xi_{0}\approx20\textrm{Å}$,
and $\mathit{j_{c}}$$\sim10^{8}A/cm^{2}$ (in cgs units $\mathit{j_{c}}$$\sim3\times10^{17}esu/cm^{2}$,
$\mathit{T_{c}}$$\sim200K$, $\Phi_{0}\approx2\times10^{-7}G\times cm^{2},$
$\mathit{L_{col}\sim20\mu m}$. As the result we obtain for $(\Delta T)/T_{c}$$\approx2\times10^{-2}$and
the ratio
\begin{equation}
\frac{(\Delta T)^{col}}{(\Delta T)^{st}}\sim5\times10^{-2}.\label{eq:Ratio}
\end{equation}

This result means, that by assuming long columnar pinning defects
the temperature broadening of the resistance is much smaller than
in standard superconductors. This is encoriging result having in mind
that the Thinkam theory might be not quite appropriate for all RTSC-hydrides.
This is demonstrated in Fig.2 where the line broadening das not depend
on the magnetic field. It might be, that the collective pinning theory
with $\mathit{a\sim(\Phi}$/$\mathit{B)^{1/2}}$ is not guaranted
in systems with long and strong pinning defects. This is a matter
for further investigation.

\section{Discussion and conclusions}

That the RTSC-hydrides are hard type-II superconductors with strongly
pinned vortices in magnetic field is clearly shown in an inventive
experiment by Troyan et al. \cite{Troyan-1}. In the setup shown schematiclly
in Fig.1 they have intended to study the Meissner effect in $\mathit{H}_{3}$$S$
hydride. It was expected that in the magnetic field $\mathit{H=0,68T}$,
which is much larger than the first critical field $\mathit{H_{c1}}$,
that the vortices penetrate into the sample, what should be detected
by the NRS technique in the non-superconducting Sn foil (yelow in
Fig.1). However, even in such a large field the vorticies do not show
up in the Sn foil. This means that they are trapped in the $\mathit{H}_{3}$$S$
sample, where the magnetization is highly inhomogeneous and hysteretic.
This is strong evidence that the pinning in $\mathit{H_{3}S}$ is
huge. By using the Bean critical state model and the fact that vortices
pentrate into the sample only the distance $\mathit{(D-d)/2}$ and
that they are pinned along some long defects with the length $\mathit{P=50\mu m}$,
one obtains that the critical current density is of the order $\mathit{\mathrm{j_{c}\sim(10^{7}-10^{8})A/cm^{2}}}$.
To this point, recently has appeared very interesting arXiv preprint
by Hirsh and Marsiglio \cite{Hirsh-Meissner}, were they have realized
that their interpretaion of the Troyan's paper as the Meissner effect
was inadequate. Namely, if it were true this would mean that the critical
current density should be enorm $\sim10^{11}A/cm^{2},$what is impossible.
Since there is a pronounced magnetization hysteresis they propose
the exsistence of strong pinning in the system. They predict much
smaller but realistic critical current density $\sim10^{7}A/cm^{2}$.

In this paper we assume the existence of long pinning defects in the
columnar form, and the pinning of vorties take place on long defects.
As the result of the strong pinning in RTSC-hydrides, the temperature
broadening of the resistance in magnetic field is drastically smaller
than the standard theory predicts. This reduction is due to the small
factor $\xi_{0}/L_{col}$. These columnar defects cause also large
magnetization hysteresis $\Delta M$ - see \cite{Troyan-1} and accordingly
high critical current density $J\sim\Delta M.$ The irreversible line
of the magnetic has the form $\mathit{H\sim(1-t)^{2}L/\xi_{0}}$,
i.e. it is drastically increased compared to the situation in HTSC-cuprates. 

There is an intriguing question - what is the physical origin for
these long defects in RTSC-hydrides? Answering on this question will
probably open a new physicsin hard type-II superconductors. 
\begin{acknowledgments}
The author would like to thank Professor Dirk Risshke for his support,
both scientific and moral, which has lasted a very long time. I am
thankful to Dr Igor Kuli\'{c} for discussions and support.
\end{acknowledgments}

\end{document}